\documentclass[]{article}
\usepackage{amssymb}
\usepackage{epsfig}
\voffset= - 0.3in
\hoffset= - 1.0in         

\catcode`\@=11
\def\marginnote#1{}

\newcount\hour
\newcount\minute
\newtoks\amorpm
\hour=\time\divide\hour by60
\minute=\time{\multiply\hour by60 \global\advance\minute by-\hour}
\edef\standardtime{{\ifnum\hour<12 \global\amorpm={am}%
        \else\global\amorpm={pm}\advance\hour by-12 \fi
        \ifnum\hour=0 \hour=12 \fi
        \number\hour:\ifnum\minute<10 0\fi\number\minute\the\amorpm}}
\edef\militarytime{\number\hour:\ifnum\minute<10 0\fi\number\minute}

\def\draftlabel#1{{\@bsphack\if@filesw {\let\thepage\relax
   \xdef\@gtempa{\write\@auxout{\string
      \newlabel{#1}{{\@currentlabel}{\thepage}}}}}\@gtempa
   \if@nobreak \ifvmode\nobreak\fi\fi\fi\@esphack}
        \gdef\@eqnlabel{#1}}
\def\@eqnlabel{}
\def\@vacuum{}
\def\draftmarginnote#1{\marginpar{\raggedright\scriptsize\tt#1}}

\def\draft{\oddsidemargin -.5truein
        \def\@oddfoot{\sl preliminary draft \hfil
        \rm\thepage\hfil\sl\today\quad\militarytime}
        \let\@evenfoot\@oddfoot \overfullrule 3pt
        \let\label=\draftlabel
        \let\marginnote=\draftmarginnote
   \def\@eqnnum{(\theequation)\rlap{\kern\marginparsep\tt\@eqnlabel}%
\global\let\@eqnlabel\@vacuum}  }

\def\appname{Appendix}
\newcounter{app}
\def\theapp{\Alph{app}}
\def\app{\par
   \addvspace{4ex}
   \@afterindentfalse
  \secdef\@app\@dapp}
\def\@app[#1]#2{\ifnum \c@secnumdepth >\m@ne
        \refstepcounter{app}
        \addcontentsline{toc}{app}{\theapp
        \hspace{1em}#1}\else
      \addcontentsline{toc}{app}{ #1}\fi
   {\parindent \z@ \raggedright
    \Large \bf \appname~\theapp .
   \Large  \bf 
    #2}\nobreak
   \vskip 4ex   \noindent
\setcounter{equation}{0}
\def\theequation{\Alph{app}.\arabic{equation}}}
\def\@dapp#1{%
{\parindent \z@ \raggedright  \bf #1}\par\nobreak}
\def\l@app#1#2{\addpenalty{\@secpenalty}%
   \addvspace{1em plus\p@}%
   \begingroup
   \@tempdima 3em
     \parindent \z@ \rightskip \@pnumwidth
     \parfillskip -\@pnumwidth
     { \bf
     \leavevmode
     #1\hfil \hbox to\@pnumwidth{\hss #2}}\par
     \nobreak
   \endgroup}

\makeatletter
\newdimen\normalarrayskip            
\newdimen\minarrayskip               
\normalarrayskip\baselineskip
\minarrayskip\jot
\newif\ifold             \oldtrue            \def\new{\oldfalse}
\def\arraymode{\ifold\relax\else\displaystyle\fi}
\def\eqnumphantom{\phantom{(\theequation)}} 
\def\@arrayskip{\ifold\baselineskip\z@\lineskip\z@
     \else
     \baselineskip\minarrayskip\lineskip1\baselineskip\fi}

\def\@arrayclassz{\ifcase \@lastchclass \@acolampacol \or
\@ampacol \or \or \or \@addamp \or
   \@acolampacol \or \@firstampfalse \@acol \fi
\edef\@preamble{\@preamble
  \ifcase \@chnum
     \hfil$\relax\arraymode\@sharp$\hfil
     \or $\relax\arraymode\@sharp$\hfil
     \or \hfil$\relax\arraymode\@sharp$\fi}}

\def\@array[#1]#2{\setbox\@arstrutbox=\hbox{\vrule
     height\arraystretch \ht\strutbox
     depth\arraystretch \dp\strutbox
width\z@}\@mkpream{#2}\edef\@preamble{\halign \noexpand\@halignto
\bgroup \tabskip\z@ \@arstrut \@preamble \tabskip\z@ \cr}%
\let\@startpbox\@@startpbox \let\@endpbox\@@endpbox
  \if #1t\vtop \else \if#1b\vbox \else \vcenter \fi\fi
  \bgroup \let\par\relax
  \let\@sharp##\let\protect\relax
  \@arrayskip\@preamble}
\def\eqnarray{\stepcounter{equation}%
              \let\@currentlabel=\theequation
              \global\@eqnswtrue
              \global\@eqcnt\z@
              \tabskip\@centering              
              \let\\=\@eqncr
              $$%
            \halign to \displaywidth  \bgroup
             \eqnumphantom \@eqnsel
      \hskip\@centering                               
    $\displaystyle  \tabskip\z@ {##}$%
    &\global\@eqcnt\@ne \hskip 2\arraycolsep
         $ \displaystyle  \arraymode{##}$\hfil
    &\global\@eqcnt\tw@ \hskip 2\arraycolsep
         $\displaystyle\tabskip\z@{##}$\hfil
         \tabskip\@centering
    &{##}\tabskip\z@\cr}
\makeatother

\newfont{\hr}{msbm10}
\newfont{\ams}{msam10}

\font\numbers=cmss12
\font\upright=cmu10 scaled\magstep1
\def\stroke{\vrule height8pt width0.4pt depth-0.1pt}
\def\topfleck{\vrule height8pt width0.5pt depth-5.9pt}
\def\botfleck{\vrule height2pt width0.5pt depth0.1pt}
\def\Zmath{\vcenter{\hbox{\numbers\rlap{\rlap{Z}\kern 0.8pt\topfleck}\kern
2.2pt
                   \rlap Z\kern 6pt\botfleck\kern 1pt}}}
\def\Qmath{\vcenter{\hbox{\upright\rlap{\rlap{Q}\kern
                   3.8pt\stroke}\phantom{Q}}}}
\def\Nmath{\vcenter{\hbox{\upright\rlap{I}\kern 1.7pt N}}}
\def\Cmath{\vcenter{\hbox{\upright\rlap{\rlap{C}\kern
                   3.8pt\stroke}\phantom{C}}}}
\def\Rmath{\vcenter{\hbox{\upright\rlap{I}\kern 1.7pt R}}}
\def\Z{\ifmmode\Zmath\else$\Zmath$\fi}
\def\Q{\ifmmode\Qmath\else$\Qmath$\fi}
\def\N{\ifmmode\Nmath\else$\Nmath$\fi}
\def\C{\ifmmode\Cmath\else$\Cmath$\fi}
\def\R{\ifmmode\Rmath\else$\Rmath$\fi}

\def\d{\partial}
\def\p{\partial}
\def\bea{\begin{eqnarray}}
\def\eea{\end{eqnarray}}

\def\beq{\begin{equation}}
\def\eeq{\end{equation}}
\def\ba{\beq\new\begin{array}{c}}
\def\ea{\end{array}\eeq}
\def\be{\ba}
\def\ee{\ea}
\def\F{{\cal F}}

\def\stackreb#1#2{\mathrel{\mathop{#2}\limits_{#1}}}
\def\Tr{{\rm Tr}}
\def\res{{\rm res}}
\def\Bf#1{\mbox{\boldmath $#1$}}

\def\bdelta{{\Bf\delta}}

\def\Im{{\rm Im}}
\def\Re{{\rm Re}}

\def\half{{\textstyle{1\over2}}}
\def\ha{{1\over 2}}
\def\N2{${\cal N}=2$}
\def\4N{${\cal N}=4$}
\def\1N{${\cal N}=1$}
\def\1N*{${\cal N}=1^*$}

\def\CS{{\cal S}}

\def\beq{\begin{equation}}
\def\eeq{\end{equation}}
\def\ba{\beq\new\begin{array}{c}}
\def\ea{\end{array}\eeq}
\def\be{\ba}
\def\ee{\ea}
\def\theequation{\thesection.\arabic{equation}}
\def\mezzo#1{\bigskip\noindent{\sl #1}\bigskip}

\def\CR{{\mathcal R}}

\newcommand{\rf}[1]{(\ref{#1})}

\begin{document}


\begin{flushright}
FIAN/TD-15/07\\
ITEP/TH-24/07
\end{flushright}
\vspace{1.0 cm}

\renewcommand{\thefootnote}{\fnsymbol{footnote}}
\begin{center}
{\Large\bf
On Microscopic Origin of Integrability in Seiberg-Witten Theory\footnote{Based
on the talks at 'Geometry and Integrability in Mathematical Physics', Moscow, May 2006;
'Quarks-2006', Repino, May 2006; Twente conference on Lie groups, December 2006
and 'Classical and Quantum Integrable Models', Dubna, January 2007.}
}\\
\vspace{1.0 cm}
{\large A.~Marshakov}\\
\vspace{0.6 cm}
{\em
Theory Department, P.N.Lebedev Physics Institute,\\
Institute of Theoretical and Experimental Physics,\\ Moscow, Russia
}\\
\vspace{0.3 cm}
{e-mail:\ \ mars@lpi.ru,\ \ mars@itep.ru}
\end{center}
\begin{quotation}
\noindent
We discuss microscopic origin of integrability in Seiberg-Witten theory, following mostly the
results of \cite{MN}, as well as present their certain extension and consider several explicit
examples. In particular, we discuss in more detail the theory with the only switched on higher
perturbation in the ultraviolet, where extra explicit formulas are obtained using bosonization
and elliptic uniformization of the spectral curve.
\end{quotation}

\renewcommand{\thefootnote}{\arabic{footnote}}
\setcounter{section}{0}
\setcounter{footnote}{0}
\setcounter{equation}0

\section{Introduction}

Supersymmetric gauge theories have become recently an area, which allows the application
of the nontrivial methods of modern mathematical physics. In particular, one is often
interested in the properties of the low-energy effective actions, which in the theories
with extended supersymmetry can be expressed in terms of the holomorphic functions
on moduli spaces of vacua, or prepotentials. These prepotentials obey remarkable
properties, which can be shortly characterized by the fact, that they are quasiclassical tau-functions \cite{KriW}, and the low-energy effective theory \cite{SW}
can be formulated in terms of an integrable system \cite{gkmmm}.

Exact form of the prepotential provides comprehensive information about the effective
theory at strong coupling, while at weak coupling the prepotential can be expanded
over the contributions of the gauge theory instantons. As often happens in conventional
quantum field theory, even each term in this infinite expansion, containing the integration over
the non-compact instanton moduli space, is ill-defined. It turns out, however, that
there exists a preserving supersymmetry infrared regularization, which allows to
perform a computation, reducing it to a sum over the point-like instantons, whose
contributions are parameterized by random partitions \cite{Nek}. Moreover, it turns our that
the regularized volume of the four-dimensional space time can be re-interpreted
as a coupling constant in dual topological string theory \cite{LMN}, providing a new form
of the gauge/string duality.

This duality predicts a nontrivial relation between the deformed
prepotentials of \N2 supersymmetric
gauge theories and the generating functions of the Gromov-Witten classes.
Similar to the latter \cite{op}, the deformed prepotentials can be expressed in
terms of the correlation functions in the theory of two-dimensional
free fermions. These correlation functions can be identified with the
tau-functions of integrable systems, whose fermionic representation is
essentially different from the conventional one \cite{UT}. We shall postpone the detailed discussion
of this issue for the full deformed prepotentials and concentrate, following \cite{MN},
their main quasiclassical asymptotic.

\section{Preliminaries}

\bigskip
\mezzo{Free fermions}

\noindent
Let us introduce, first, the main definitions and notations for
the two-dimensional theory of a single free complex fermion the action $\int
{\tilde\psi} {\bar\d} {\psi}$ on a cylinder.
One can expand the solutions to Dirac equation in holomorphic co-ordinate $w\in\mathbb{C}^\ast$:
\be
\label{frem}
{\psi} (w) = \sum_{r \in \mathbb{Z} + {\ha}} \
{\psi}_{r} \  w^{-r} \left( dw \over w \right)^{\half},
\\
{\widetilde\psi} (w) = \sum_{r \in \mathbb{Z} + {\ha}} {\widetilde
\psi}_{r} \ w^r \left( dw \over w \right)^{\half},
\ee
so that the modes after quantization satisfy the (anti)commutational relations
\be
\{ {\psi}_{r}, {\widetilde\psi}_{s} \} = {\delta}_{rs}
\ee
The fermionic Fock space is constructed with the help of the
charge $M$ vacuum state (a Dirac sea)
\be
\label{vcms}
\vert M \rangle = {\psi}_{-M + {1\over
2}} {\psi}_{-M+{3\over 2}} {\psi}_{-M+ {5\over 2}} \ldots =
\bigwedge_{r>-M}\psi_r
\ee
with
\be
{\psi}_{r} \vert M \rangle = 0,\ r > -M,\ \ \ \ \
 {\widetilde\psi}_{r} \vert M \rangle = 0,\ r < -M
\ee
and these definitions correspond to the two-point function
\be
\label{vaccor}
\langle 0 |{\tilde\psi} (z)\psi (w)| 0\rangle = {\sqrt{dz dw}\over z-w}
\ee
More conventional ``Japanese" conventions (with the integer-valued fermionic operators
$\psi_i$, $\psi^*_i$, $i\in \mathbb{Z}$, see e.g. \cite{JM}) can be got from these by
\be
\label{japfer}
{\tilde\psi}_r \to \psi_{r+\half},\ \ \ \
\psi_r \to \psi^\ast_{r+\half},\ \ \ \
M = - n
\ee
It is also convenient to use the basis of the so-called
partition states: for each partition ${\bf k} = ( k_1  \geq
k_2 \geq \ldots \geq k_{\ell_{\bf k}}=0 \geq 0 \ldots)$ one introduces the state:
\be
\label{prtst}
\vert M;
{\bf k} \rangle = {\psi}_{-M+{1\over 2} -k_1} {\psi}_{-M+{3\over 2} -
k_2} \ldots =
\bigwedge_{r>-M}\psi_{r-k_i}
\ee
and defines the $U(1)$ current as:
\be \label{crnt}
J = : {\widetilde\psi} {\psi} : = \sum_{n \in \mathbb{Z}} J_{n} w^{-n} {dw
\over w}, \ \ \ \ \
J_{n} = \sum_{r\in \mathbb{Z} + {\ha}} :{\widetilde\psi}_{r}{\psi}_{r+n}:
\ee
Obviously
\be
\label{Jpsi}
\left[ J_n,\psi_r\right]=-\psi_{r+n},\ \ \ \
\left[ J_n,{\tilde\psi}_r\right]={\tilde\psi}_{r-n}
\\
\left[ J_n,\psi(w)\right]=-w^n\psi(w),\ \ \ \
\left[ J_n,{\tilde\psi}(w)\right]=w^n{\tilde\psi}(w)
\ee
Recall the bosonization rules:
\be \label{bsnz}
{\tilde\psi} = : e^{i {\phi}}
:\ , \quad {\psi} = : e^{-i{\phi}} :\ , \quad J =
i{\p} {\phi}
\ee
where
\be
\label{fiope}
\phi (z)\phi (0) \sim - \log z + \ldots
\ee
and a useful fact from $U({\hat N})$ and permutation's group theory:
the Schur-Weyl correspondence, which states that
\be \label{wlcrsp}
({\bf C}^{{\hat N}})^{\otimes k} = \bigoplus_{{\bf k}, \vert
{\bf k} \vert = k} R_{\bf k} \otimes {\CR}_{\bf k}
\ee
as ${\CS}_{k} \times
U({\hat N})$ representation. Now let $U = {\rm diag} \left( u_1,
\ldots, u_{\hat N} \right)$ be a $U({\hat N})$ matrix. Then one
easily gets using the Weyl character formula, and the bosonization
rules \rf{bsnz}, that:
\be
\label{chrt}
{\Tr}_{{\CR}_{\bf k}} U =
\langle {\hat N}; {\bf k} \vert : e^{i \sum_{n=1}^{{\hat N}} {\phi}
( u_n) } : \vert 0 \rangle \ = \ s_{\bf k}(u_1,\ldots,u_{\hat N}) =
{\det u_j^{k_i+{\hat N}-i}\over \det u_j^{{\hat N}-i}}
\ee
gives the (ratio of the) standard Schur functions for any partition ${\bf k}$,
a very nice review of their properties can be found in \cite{Olsh}. In particular,
from this formula one derives:
\be \label{nashevse}
e^{J_{-1}\over {\hbar}} \ \vert M \rangle
= \sum_{\bf k} {{\bf m}_{\bf k} \over {\hbar}^k}\ \vert M; {\bf k} \rangle
= \sum_{\bf k}
{{\rm dim}R_{\bf k} \over {\hbar}^k \ k! }\ \vert M; {\bf k} \rangle
\ee
with
\be
\label{planch}
{\bf m}_{\bf k} = {{\rm dim}R_{\bf k} \over \ k! } = \prod_{i<j}{k_i-k_j+j-i\over j-i}=
\\
=  \prod_{i=1}^{\ell_{\bf k}}{(\ell_{\bf k}-i)!\over(\ell_{\bf k}+k_i-i)!}
\prod_{1\leq i<j \leq \ell_{\bf k}}{k_i-k_j+j-i\over j-i}
= {\prod_{1\leq i<j \leq \ell_{\bf k}}(k_i-k_j+j-i)\over
\prod_{i=1}^{\ell_{\bf k}}(\ell_{\bf k}+k_i-i)!}
\ee
being the Plancherel measure. It follows from the fact that for particular
values $u_1=\ldots=u_{\hat N}={1\over\hbar{\hat N}}$
\be
\label{schupla}
s_{\bf k}\left({1\over\hbar{\hat N}},\ldots,{1\over\hbar{\hat N}}\right)\
\stackreb{{\hat N}\to\infty}{=}\ {{\bf m}_{\bf k}\over \hbar^k}
\ee

\bigskip
\mezzo{Instantons and Nekrasov's computation}

\noindent
In the context of \N2 supersymmetric gauge theories, one usually
starts with the microscopic theory, determined by the
ultraviolet prepotential ${\F}_{\rm UV}$, which can be taken perturbed by arbitrary powers of the
holomorphic operators
\be
\label{fuv}
{\F}_{\rm  UV} = {\half} ( {\tau}_{0} +  t_{1})\  {\Tr}\ {\Phi}^{2} +
\sum_{k > 0} t_{k}\ {{\Tr}\ {\Phi}^{k+1}\over k+1} \equiv {\half}{\tau}_{0}{\Tr}\ {\Phi}^{2}
+ \Tr\ {\bf t}(\Phi)
\ee
and quadratic ${\overline{\F}_{\rm UV}}= {\half}{\bar\tau}_{0}\ {\Tr}\ {\bar \Phi}^{2}$.
Then one integrates out the fast modes, i.e. the perturbative
fluctuations with momenta above certain scale ${\mu}$ as well as the non-perturbative modes, e.g. instantons
(and fluctuations around them) of all sizes smaller
then ${\mu}^{-1}$. The resulting effective theory has a derivative expansion
in the powers ${\p}^{2}\over {\mu}^{2}$. The leading terms in the expansion are all determined,
thanks to the \N2 supersymmetry,
by the effective prepotential ${\F}({\mu})$. As ${\mu}$ is lowered all the
way down to zero, we arrive at the infrared prepotenial ${\F}_{\rm UV} \rightarrow
{\F}_{\rm IR}$.
The supersymmetry considerations suggest that the renormalization flows of
${\F}$ and ${\bar\F}$ proceed more or less independently from each other.
Thus one can simplify the problem by taking the limit, ${\bar\tau}_{0} \to i \infty$,
while ${\F}_{\rm UV}$ kept fixed. In this limit the
path integral is dominated by the gauge instantons. The setup of \cite{Nek} allows to
evaluate their contribution, as well as the contribution of the fluctuations around
the instantons, exactly. The price one pays is the introduction of extra parameters
into the problem, some sort of the infrared cutoff, which we denote
by ${\hbar}^{-2}$, since it appears to be a parameter of the loop expansion
in dual topological string theory \cite{LMN}.

In particular, for the so-called noncommutative $U(1)$ theory, or the theory
on a single D3 brane in the background, which preserves only sixteen supercharges (so that
the theory on the brane has only eight supercharges)\footnote{This
theory can also be realized at a special point on the moduli space of $U(N)$ gauge
theory with $2N-2$ fundamental hypermultiplets.},
the instanton partition function $Z(a, {\hbar}, {\bf t})$,
${\bf t} = \left( t_{1} , t_{2} , \ldots \right)$, can be shown to be given by
the sum over the Young diagrams, i.e. over the partitions \cite{Nek,LMN,NO}:
\be\label{zuone}
Z ( a, {\bf t}, {\hbar} ) =
\sum_{{\bf k}}
{{\bf m}_{\bf k}^2\over{( - {\hbar}^2)^{| {\bf k} |}}} \
{\exp} \ {1\over {\hbar}^{2}} \sum_{k>0}
t_{k}{{\rm ch}_{k+1} ( a, {\bf k},\hbar)\over k+1}
\ee
where ${\bf m}_{\bf k}$ is the Plancherel measure \rf{planch}, and the Chern
polynomials ${\rm ch}_{k+1} ( a, {\bf k},\hbar)$ can be introduced, e.g. via
\be
\label{chern}
\left( e^{{\hbar}u\over 2} -
e^{-{\hbar u \over 2}} \right)
\sum_{i=1}^{\infty} e^{ u ( a + \hbar({\half} - i +k_{i}))}=
 \sum_{l=0}^{\infty}\
{u^l\over l!}\ {\rm ch}_{l}( a, {\bf k},\hbar)
\ee
If the theory has the gauge group $U(N)$, e.g. it is realized on the
stack of $N$ fractional D3 branes, the corresponding partition function is given by
the generalization of \rf{zuone}:
\be
\label{zunntre}
Z ( {\vec a},  {\bf t}, {\hbar} ) =  Z^{\rm pert} ( {\vec a} , {\bf t}, {\hbar}  )
\sum_{{\vec {\bf k}}} \left(  {\bf m}
( {\vec a}, {\vec{ \bf k}}, {\hbar})   \right)^{2} \ (-1)^{| {\vec{\bf k}}|}
{\exp} \ {1\over {\hbar}^{2}} \sum_{k>0} t_{k}
{{\rm ch}_{k+1}( {\vec a}, {\vec{\bf k}},\hbar)\over k+1}
\ee
where ${\bf m}( {\vec a}, {\vec {\bf k}}, {\hbar})$ is the $U(N)$ generalization of
Plancherel measure \cite{NO} and $ Z^{\rm pert} ( {\vec a} , {\bf t}, {\hbar}  ) $
is the perturbative partition function.

\bigskip
\mezzo{Toda chain and tau-functions}

\noindent
Consider, first, the well-known formula for the tau-function of Toda molecule (or the open N-Toda
chain with co-ordinates
$q_n({\bf t},{\sf a})=\log{Z({\bf t};n| {\sf a})\over Z({\bf t};n-1| {\sf a})}$ \cite{Toda}),
given by all principal $n$-minors
\be
\label{tauto}
Z({\bf t}; n | {\sf a}) =
 \sum_{K: i_1<\dots <i_n}\mu_{K}({\sf a})^2
\exp \sum_{l,i_k}t_lH_l({\sf a}_{i_k}) =
 \Delta_{n\times n}\left(A\cdot D\cdot A^T\right)
\ee
of the $N\times N$ matrix, expressed as a matrix product
with $A_{ij} \sim {\sf a}_i^{j-1}$, $D_{ij}=
\delta_{ij}\exp(z_i)\prod^N_{k=1,k\neq i}|{\sf a}_i-{\sf a}_k|^{-1}$, where
\be\label{angles}
z_i=\sum_{l}t_lH_l({\sf a}_i)=\sum_{l}(t_l{\sf a}_{i}^l+\ldots)+z_i^{(0)}
\ee
with some appropriately chosen "initial phases"
$z_i^{(0)}$. Rewriting \rf{tauto} in the form
\be
\label{brama}
Z({\bf t}; n| {\sf a})
={\sum}_{|K|=n}\prod_{i\in K}{e^{z_i}\over
\prod^N_{k=1,k\neq i}|{\sf a}_i-{\sf a}_k|}\prod_{i,j\in K,i\neq j} ({\sf a}_i-{\sf a}_j)^2
\ee
we see that the sum in \rf{tauto} is in fact taken over the
partitions $(k_1 \geq k_2 \geq \ldots \geq k_n)$ with the fixed length and
\be
\label{jk}
k_j = i_{n-j+1}+j-n-1,\ \ \ j=1,\ldots,n
\ee
For the particular solution of the Toda chain with ${\sf a}_i\simeq i$, one gets for \rf{brama}
\be
\label{toi}
Z({\bf t}; n)
={\sum}_{|K|=n}\prod_{i\in K}e^{z_i}{\prod_{i,j\in K,i\neq j} (i-j)^2\over
\prod_{i\in K}(i-1)!(N-i)!}
\ee
This is a singular or "stringy'' solution, presenting a collection of particles, moving
each with a constant speed, proportional to its number. In KP/KdV-theory the analog is
$u \propto x/t$, a linear growing potential of Kontsevich model \cite{GKM}, which never topples.

Comparing \rf{toi} to \rf{tauto}, one finds that
\be
\label{mum}
\left.\mu_{K}({\sf a})^2\right|_{{\sf a}_i=i} = \left({\prod_{i,j\in K} (i-j)\over
\prod_{i\in K}(i-1)!}\right)^2\prod_{i\in K}{(i-1)!\over(N-i)!}\
= \left.{\bf m}_{\bf k}^2\right|_{\ell_{\bf k}=n}\ \prod_{i\in K}e^{z_i^{(0)}}
\ee
In the limit $N\to\infty$, after particular choice of the Hamiltonians \rf{chern}
\be\label{hamchern}
\left.H_l({\sf a}_i)\right|_{{\sf a}_i=i} \rightarrow {{\rm ch}_{l+1}(a ; i, \hbar)\over l+1}
\ee
and renormalization of the initial phase $z_i^{(0)}$, passing
from summation over partitions with a fixed length $\ell_{\bf k}=n$ to a "grand-canonical'' ensemble
by a sort of Fourier transform, one gets
$Z({\bf t}; n)\ \rightarrow\ Z ( a , {\bf t}, {\hbar} )$ the \rf{zuone} partition function.
By \rf{nashevse}, it becomes
equivalent to the following fermionic correlator
\be
\label{zumat}
Z ( a , {\bf t}, {\hbar} )  =
\sum_{{\bf k}}
{{\bf m}_{\bf k}^2\over{( - {\hbar}^2)^{| {\bf k} |}}} \
e^{ {1\over {\hbar}^{2}} \sum_{k>0}
t_{k}{{\rm ch}_{k+1} ( a, {\bf k})\over k+1}}=
\langle M \vert e^{-{J_{1}\over \hbar}} e^{{1\over\hbar}\sum_{k>0}
t_{k} {W}_{k+1}} e^{J_{-1} \over \hbar} \vert M \rangle
\ee
where the mutually commuting modes of the $W$-infinity generators can be defined as
\be
\label{wgen}
W_{k+1} =
-{{\hbar}^{k}\over k+1}
\oint : {\tilde\psi} \left( \left(w{d\over dw} + {1\over 2}\right)^{k+1}  -
\left(w{d\over dw} - {1\over 2} \right)^{k+1} \right) {\psi} :\ =
\\
= {{\hbar}^{k} \over k+1}
\sum_{r \in \mathbb{Z} +{\half}}   \left[  (- r+{\half})^{k+1} -
(- r - {\half} )^{k+1} \right] \ : {\psi}_{r}{\tilde\psi}_{r} :
\ee
The matrix element \rf{zumat} is a particular non-standard fermionic representation of
the tau-function, where the Toda times are coupled to the $W$-generators \rf{wgen} instead
of the modes of the $U(1)$ current \rf{crnt}, and it has been discussed in \cite{op,OW}.

If only $t_1
\neq 0$ the correlator in \rf{zumat} gives
\be
\label{taucp1}
Z(a=\hbar M,t_1,0,0,\ldots) = \langle M \vert e^{-{J_{1}\over \hbar}} e^{{1\over\hbar}
t_{1} L_0 } e^{J_{-1} \over \hbar} \vert M \rangle
 = \exp\left[ -{1\over {\hbar}^2} \left( \half t_1 a^2
 + e^{t_1} \right)\right]
= \exp \left(-{\F_{\mathbb{P}^1}\over {\hbar}^2}\right)
\ee
the partition function of topological string on $\mathbb{P}^1$. This is the only case when summing over partitions can be performed straightforwardly, using the Burnside theorem
\be
\label{burn}
\sum_{{\bf k}, |{\bf k}|=k} {\bf m}_{\bf k}^2 =
{1\over \ k!^2 }\sum_{{\bf k}, |{\bf k}|=k}
{\rm dim}R_{\bf k}^2 = {1\over \ k! }
\ee

\newpage
\mezzo{Baker-Akhiezer functions}

\noindent
In addition to \rf{nashevse}, one can consider
\be
\label{tpsiact}
\tilde{\psi}_{-r}\ e^{J_{-1}\over \hbar} \vert M+1 ; {\emptyset}
\rangle
= \sum_{\bf k} {\tilde C}_{\bf k} \vert M ; {\bf k} \rangle
\ee
with (computed by the Wick theorem and using the properties of the Schur
functions \rf{chrt})
\be
\label{nashekhar}
{\tilde C}_{\bf k} = \langle M, {\bf k} |\tilde{\psi}_{-r}\ e^{J_{-1}\over \hbar}
\vert M+1 ; {\emptyset}
\rangle = \hbar^{M-|{\bf k}|-r-\half}\prod_{i=1}^{\infty}
{ i - k_{i} + r-\half-M \over i }\ {\bf m}_{\bf k}
\ee
where the infinite product is actually finite
\be\label{infprd}
\prod_{i=1}^{\infty}
{ i - k_{i} + r-\half-M\over i }   = {1\over \Gamma(r+\half-M)}
\prod_{i=1}^{{\ell}_{\bf k}}
{ i - k_{i} + r-\half-M\over i +r-\half-M}
\ee
Therefore,
one gets for the Baker-Akhiezer functions
\be
\label{batpsir}
{\tilde\Psi} (r) = {\langle M \vert  e^{-{J_{1}\over \hbar}} {\tilde\psi}_{-r}
e^{{1\over\hbar}\sum_{k>0} t_{k} {W}_{k+1}}
e^{J_{-1} \over \hbar} \vert M+1 \rangle\over
\langle M \vert  e^{-{J_{1}\over \hbar}}
e^{{1\over\hbar}\sum_{k>0} t_{k} {W}_{k+1}}
e^{J_{-1} \over \hbar} \vert M \rangle} =
\\
= {\hbar^{M-r-\half}\over Z(a,{\bf t},\hbar)}
e^{{1\over\hbar^2}\sum_{k>0} {t_{k}\hbar^k\over k+1}
\left((r+\half)^{k+1}-(r-\half)^{k+1}\right)}\sum_{\bf k}
{{\bf m}_{\bf k}^2\over{( - {\hbar}^2)^{| {\bf k} |}}} \
e^{{1\over {\hbar}^{2}} \sum_{k>0}
t_{k}{{\rm ch}_{k+1} ( a, {\bf k},\hbar)\over k+1}}
\prod_{i=1}^{\infty}
{i-k_{i}+r-\half-M\over i } =
\\
= \hbar^{M-r-\half}\exp{1\over\hbar^2}\sum_{k>0} {t_{k}\hbar^k\over k+1}
\left((r+\half)^{k+1}-(r-\half)^{k+1}\right)
\cdot{e^{M\Gamma'(r+\half)/\Gamma(r+\half)}\over\Gamma(r+\half)}\cdot
{Z (a, {\bf t} - {\bdelta}(r),\hbar) \over
Z ( a, {\bf t},\hbar)}
\ee
with $a=M\hbar$ and the shift $\bdelta(r)=(\delta_1(r),\delta_2(r),\ldots)$ generated by
\be\label{dellog}
{1\over \hbar^2}\sum_{k>0}  {\delta}_{k}(r) {x^{k+1}\over k+1}
= {\rm log}  {{\Gamma} \left(  r +\half- { x\over\hbar}\right) \over
{\Gamma} \left(  r + \half \right) }
+ {x\over\hbar}\ {\Gamma'(r+\half)\over\Gamma(r+\half)}
\ee
In the quasiclassical asymptotic $\hbar\to 0$, with $\hbar r=z$, \rf{batpsir} gives
\be
\label{quasf}
{\tilde\Psi}(z, a, {\bf t}, {\hbar}) \sim \exp{1\over\hbar}\left(
\sum_{k>0}t_kz^k - z(\log z-1) + a\log z + \ldots\right)
\ee
(similar formulas in the context of five-dimensional Seiberg-Witten theory \cite{5dSW} were
considered in \cite{5dTak}).
In the same way one can define the two-point function
\be
\label{nashepp}
{\cal E}(r) = {\langle M+1 \vert  e^{-{J_{1}\over \hbar}} \psi_{-r}{\tilde\psi}_{-r}
e^{{1\over\hbar}\sum_{k>0} t_{k} {W}_{k+1}}
e^{J_{-1} \over \hbar} \vert M+1 \rangle\over
\langle M \vert  e^{-{J_{1}\over \hbar}}
e^{{1\over\hbar}\sum_{k>0} t_{k} {W}_{k+1}}
e^{J_{-1} \over \hbar} \vert M \rangle}\sim
\\
\sim\exp{1\over\hbar^2}\sum_{k>0} {t_{k}\hbar^k\over k+1}
\left((r+\half)^{k+1}-(r-\half)^{k+1}\right)
\cdot{e^{2M\Gamma'(r+\half)/\Gamma(r+\half)}\over\Gamma(r+\half)^2}\cdot
{Z (a, {\bf t} - 2\cdot{\bdelta}(r),\hbar) \over
Z ( a, {\bf t},\hbar)}\sim
\\
\stackreb{\hbar\to 0}{\sim}\ \exp{S(z, a, {\bf t}) \over \hbar}
\ee
with
\be
\label{Sas}
S(z, a, {\bf t}) = \sum_{k>0}t_kz^k - 2z(\log z-1) + 2a\log z + \ldots
\ee
The asymptotics \rf{Sas} plays an essential role in the study of the
quasiclassical solution. Formula \rf{nashepp} can be also interpreted
as average of the $r$-th Fourier mode of the ``symmetrically splitted'' bi-fermionic operator
${\cal E}(\zeta) = \oint_{dw\over w}
\psi\left(we^{-\zeta/2}\right){\tilde\psi}\left(we^{\zeta/2}\right)$,
introduced in \cite{op}. The ``doubling'' of the fermions and their
symmetric splitting along the $w$-cylinder turn into the double covering
of $z$-plane by the quasiclassical spectral curve.

\bigskip
\mezzo{Bosonization}

\noindent
On a small phase space ($t_k=0$ with $k>1$) the tau-function of $\mathbb{P}^1$ model \rf{taucp1}
can be easily computed exploiting the fact that it can be presented as a matrix element of the evolution operator in
harmonic oscillator for the initial and final coherent states.
Indeed, after identifying
$L_0=\half J_0^2+J_{-1}J_1=-\half M^2+{1\over\hbar^2}\alpha^\dagger \alpha$, where $[\alpha,\alpha^\dagger]=\hbar^2$, on the eigenstates with $J_0\sim M={a\over\hbar}$, one gets for \rf{zumat}
\be
Z(a,t_1,0,0,\ldots)=\exp\left(-{ t_1a^2\over 2\hbar^2}\right)\langle 0 |e^{-{\alpha\over\hbar^2}}
e^{{t_1\over\hbar^2} \alpha^\dagger \alpha}
e^{{\alpha^\dagger\over\hbar^2}}|0\rangle
= \exp\left(-{ t_1a^2\over 2\hbar^2}\right)\sum_{n\geq 0}{e^{t_1n}\over (-\hbar^2)^n n!} =
\\
= \exp\left[-{1\over\hbar^2}\left(\half t_1a^2+
e^{t_1}\right)\right]
\ee
where independent of the charge $a$ part (generated by the world-sheet instantons in $\mathbb{P}^1$ topological
string model) is just a kernel of the evolution operator in the holomorphic representation
with fixed at the boundaries $\alpha_{\rm in}$ and $\alpha^\dagger_{\rm out}$. This result is
certainly exact quasiclassically, here in the ``stringy normalized'' Planck constant ($\hbar^2$ instead of $\hbar$).

If the second time $t_2$ is also switched on, the partition function
\be
Z(M;t_1,t_2,0,0,\ldots)=\langle M |e^{-{J_1\over\hbar}}e^{t_1 L_0 + \hbar t_2W_0}
e^{{J_{-1}\over\hbar}}|M\rangle
\ee
is no longer described in terms of a single quasiparticle. Now one gets
\be
\label{l0}
L_0 \rightarrow H_0 = \sum_{n>0}n\alpha^\dagger_n\alpha_n = \alpha^\dagger \alpha + 2A^\dagger A
+\ldots
\ee
the system of coupled oscillators $[\alpha_n,\alpha^\dagger_m]=\hbar^2\delta_{nm}$
with the quadratic Hamiltonian,
perturbed by special, preserving the energy due to $[H_0,H_I]=0$, interaction
\be
\label{w0}
W_0 \rightarrow H_I=\sqrt{2}\left((\alpha^\dagger)^2A+\alpha^2A^\dagger\right) + \ldots
\ee
More strictly,
\be
\label{Zt2}
Z(M;t_1,t_2,0,0,\ldots)=e^{{\bf t}(a)\over\hbar^2}
\langle 0 |e^{-{\alpha\over\hbar^2}}\exp{1\over\hbar^2}\left( {\bf t}''(a)H_0 + t_2H_I\right)
e^{\alpha^\dagger\over\hbar^2}|0\rangle
\ee
with ${\bf t}(a) = {t_1 a^2\over 2}+{t_2 a^3\over 3}$, ${\bf t}''(a)\equiv t_1+t_2a$. Therefore
the problem reduces to the computation of the matrix elements
\be
\label{mael}
\langle 0 |e^{-{\alpha\over\hbar^2}}\exp{1\over\hbar^2}\left( {\bf t}''(a)H_0 + t_2H_I\right)
e^{\alpha^\dagger\over\hbar^2}|0\rangle =
\sum_{k=0}^\infty
{t_2^{2k}\over (\hbar^2)^{2k}(2k)!}\langle \lambda| H_I^{2k} |\lambda'\rangle =
\ee
for the coherent states
\be
\label{costa}
\alpha_n|\lambda\rangle = \lambda\delta_{n,1}|\lambda\rangle
\ee
with $\lambda=-{1\over\hbar^2}$ and $\lambda'={1\over\hbar^2}\exp {\bf t}''(a)$,
$\langle\lambda|\lambda'\rangle = \exp\left(- {{\bf t}''(a)\over\hbar^2}\right)$.

\begin{figure}[tp]
\epsfysize=2.5cm
\centerline{\epsfbox{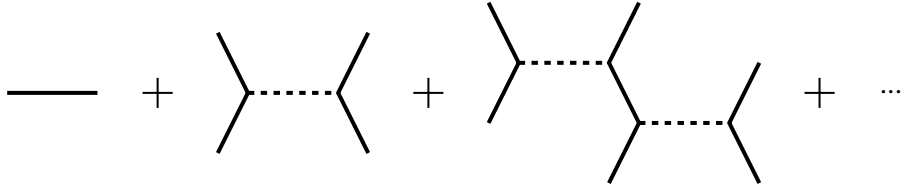}}
\caption{Prepotential, as a sum of all connected tree diagrams in the bosonic cubic field
theory. Each vertex is weighted by $t_2$ and each pair of external legs -- by $e^{{\bf t}''(a)}$.
The depicted diagramms correspond literally to the contribution of truncated ``bosonic BCS model'',
with the dashed lines being the $\langle AA^\dagger\rangle$-``propagators''. }
\label{fi:bks}
\end{figure}

Quasiclassically, the matrix element \rf{mael} gives
\be
\label{pathint}
\langle 0 |e^{-{\alpha\over\hbar^2}}\exp{1\over\hbar^2}\left( {\bf t}''(a)H_0 + t_2H_I\right)
e^{\alpha^\dagger\over\hbar^2}|0\rangle
\stackreb{\hbar\to 0}{\sim}\ \exp\left(-{1\over\hbar^2}{\sf F}({\bf t}'',t_2)\right)
\ee
so that for the prepotential $\log Z = -{1\over\hbar^2}\F + O(1)$ one gets from \rf{Zt2},\rf{pathint}
\be
\label{prelam}
{\cal F} = {\bf t}(a) + {\sf F}({\bf t}''(a),t_2)
\ee
Its nontrivial part ${\sf F}({\bf t}''(a),t_2)$ can be presented as a sum over all
connected tree diagrams (see fig.~\ref{fi:bks})
in the bosonic cubic field theory \rf{mael}, which is
encoded by appearance of the Lambert function in the exact quasiclassical solution. An interesting
issue would be to solve this theory exactly, at least quasiclassically, which is already not quite
obvious even for the ``truncated BCS model'' of two coupled oscillators, corresponding to
the first terms in \rf{l0}, \rf{w0} and diagrams depicted at fig.~\ref{fi:bks}. The corresponding
classical system
\be
\label{classpr}
\dot{\alpha} = {\bf t}''\alpha + 2t_2\alpha^\dagger A,
\ \ \ \ \
\dot{\alpha}^\dagger = -{\bf t}''\alpha^\dagger - 2t_2\alpha A^\dagger
\\
\dot{A} = 2{\bf t}''A + t_2\alpha^2,
\ \ \ \ \
\dot{A}^\dagger = -2{\bf t}''A^\dagger - t_2{\alpha^\dagger}^2
\ee
can be integrated in terms of elliptic functions, and possesses, in particular,
a "kink" solution.
We shall return to a detailed discussion of these issues elsewhere.

\section{Quasiclassical free energy}

The instantonic calculation in \N2 extended gauge theory \rf{fuv} gives rise to the Seiberg-Witten
prepotential as a critical value of the functional\footnote{We choose here different (by a factor of $2$) normalization of the time variables ${\bf t}$ compare to \cite{MN}
and correct some misprints.}
\be
\label{functnl}
\F = {1\over 2}\int dx f''(x)\sum_{k>0}t_k{x^{k+1}\over k+1} -
{1\over 2}\int_{x_1>x_2} dx_1 dx_2 f''(x_1)f''(x_2) F(x_1-x_2)
\ee
extremized w.r.t. second derivative of the profile function $f''(x)={d^2 f\over dx^2}$ with the kernel
\be
\label{SWkern}
F(x)={x^2\over 2}\left(\log x -{3\over 2}\right)
\ee
coincides with the perturbative prepotential of pure \N2 supersymmetric Yang-Mills theory.
Formula \rf{functnl} means that the quasiclassical free energy for the
partition functions \rf{zuone} and \rf{zunntre} is saturated onto a single
``large" partition ${\bf k}^\ast$ with the profile function $f_{{\bf k}^\ast}(x)=
f(x)$, where for each partition ${\bf k}=k_1\geq k_2\geq\ldots\geq k_{\ell_{\bf k}}\geq
k_{\ell_{\bf k}+1}=0,\ldots$ the profile function is defined by
\be
\label{fpart}
f_{\bf k}(x) = |x-a| + \sum_{i=1}^{\ell_{\bf k}}\left(
|x-a-{\hbar}({k}_{i}-i +1)|-|x-a -{\hbar}({k}_{i}-i)|
-|x-a-{\hbar}(1-i)|+|x-a+{\hbar}i|\right)
\ee
(see \cite{NO,MN} for details). In particular, one can write for \rf{chern}
\be
\label{chernf}
{\rm ch}_{l}(a, {\bf k}) = \ha \int \ dx \ f_{\bf k}''(x) x^l
\sim \sum_{i=1}^{\infty}
\left( (a + {\hbar} ( k_{i} - i +1 ))^l - (a +{\hbar}
( k_{i} - i  ))^l\right)
\ee
The variational problem for the functional
\rf{functnl} should be solved upon normalization condition for $f(x)$ and the constraint
\be
a = \half\int dx\ xf''(x)
\ee
which can be in standard way taken into account by adding it with the
Lagrange multiplier
\be
\label{Lagr}
\F \rightarrow \F + a^D\left(a-\half\int dx\ xf''(x)\right)
\ee
having a sense of the $k=0$ term in the summation in formula \rf{functnl}. The whole
setup of \rf{functnl} is almost identical to the standard quasiclassics of the
matrix models, where the Coulomb gas kernel is replaces by a (multivalued!) Seiberg-Witten
function \rf{SWkern}.

The extremal equation for the
\rf{functnl} gives
\be
\label{exeq}
\sum_{k>0} t_k x^k - \int d{\tilde x} f''({\tilde x})(x-{\tilde x})
\left(\log |x-{\tilde x}|-1\right) = a^D
\ee
on the support ${\bf I}$ where $f''(x)\neq 0$. Generally, for the microscopic non-abelian theory
this support consists of a set of several (disjoint) segments along the real axis
in the complex plane, where the filling fractions are fixed separately with the help of
several Lagrange multipliers, see below.
Equation \rf{exeq} means that
\be
\label{Sfun}
S(z) = \sum_{k>0}t_k z^k - \int dx f''(x)(z-x)\left(\log(z-x)-1\right)-a^D
\ee
is an analytic multivalued function on the double-cover of the $z$-plane with
the following properties:
\begin{itemize}
  \item  The real part
  of the multivalued function \rf{Sfun} vanishes
\be
\label{Scut}
S(x) = \half\left(S(x+i0)+S(x-i0)\right) =  0,\ \ \ \ x\in {\bf I}
\ee
  on the cut, due to \rf{exeq}.
  \item For its imaginary part one can write
\be
\label{ImS1}
{1\over\pi}\Im\; S(z\pm i0) = \mp \int_z^\infty dx f''(x)(z-x)
= \mp \left\{
\begin{array}{c}
  0\ ,\ \ \ \ z>x^+ \\
  a-z+f(z)\ ,\ \ \ x^-<z<x^+ \\
  2(a-z)\ ,\ \ \ z<x^-
\end{array}\right.
\ee
  \item We see from \rf{ImS1} that even the differential $dS$ is multivalued. Indeed, one
  can easily establish for
\be
\label{fi}
\Phi= {dS\over dz} = {\bf t}''(z) - \int dx f''(x)\log(z-x)
\ee
that
\be
\label{Imfi}
{1\over\pi}\Im\; \Phi(z\pm i0)
= \pm \left\{
\begin{array}{c}
  0\ ,\ \ \ \ z>x^+ \\
  1-f'(z)\ ,\ \ \ x^-<z<x^+ \\
  2\ ,\ \ \ z<x^-
\end{array}\right.
\ee
  However, the differential
\be\label{dfi}
  d\Phi = {\bf t}'''(z) dz + dz\int {dx f''(x)\over z-x}
\ee
is already single-valued on the double
  cover of the cut $z$-plane with the periods $\oint d\Phi \sim 4\pi i\mathbb{Z}$,
  so $dS$ is defined modulo $4\pi i dz$, and
  one can make sense of the periods $\oint dS$ due to $\oint dz=0$. Therefore, the exponent
  $\exp\left(\Phi/2\right)$ is already single-valued
  on the double cover and equals to unity on the cut.
  \item In order to consider the asymptotic of \rf{Sfun} in what follows we shall always
  choose a branch, which is real along the real axis, i.e. take it at
  {\em real} $x\to +\infty$. In particular, all residues below could be understood
  in this sense, as coefficients of expansion of generally multivalued
  differential at $x\to +\infty$.
  \item Taking derivatives of \rf{exeq} in $x$-variable, or integrating by parts, one
  can bring it literally to the form, arising in the context of matrix model. However,
  for the purposes of Seiberg-Witten theory one needs a solution with different analytic
  properties: in matrix models the resolvent $G \sim {dS\over dz}$ does not have poles at
  the branching points where $dz=0$ (see e.g. \cite{mamo1} and references therein), which is not
  true for \rf{Sfun}.
\end{itemize}

Asymptotically from \rf{Sfun} one gets
\be
\label{sasympt}
S(z)\ \stackreb{z\to\infty}{=}\ - 2z(\log z-1) + \sum_{k>0}t_k z^k + \log z\int dx xf''(x)  - a^D
- 2\sum_{k=1}^\infty {1\over kz^k}\int dx f''(x){x^{k+1}\over k+1} =
\\
= - 2z(\log z-1) + \sum_{k>0}t_k z^k + 2a\log z - {\d\F\over\d a}
- 2\sum_{k=1}^\infty {1\over kz^k}{\d\F\over\d t_k}
\ee
where, according to \rf{functnl}, (with convention that $\res_\infty {dz\over z}=1$)
\be
\label{derF}
{\d\F\over\d t_k} = {1\over 2(k+1)}\int dx f''(x) x^{k+1},\ \ \ \ k>0
\ee
and, due to \rf{Lagr}
\be
\label{aad}
a^D = {\d\F\over \d a}
\ee
The coefficient at the $z(\log z -1)$ term is fixed by normalization
\be
\label{normlog}
\int_{\bf I} dx f''(x) = f'(x^+)-f'(x^-)=2
\ee
where $x^\pm$ (in the one-cut case) can be defined as two solutions to the equation
\be
\label{eqends}
f(x^\pm)=|x^\pm-a|
\ee
Using variational equation \rf{exeq}, one can also write for the
functional \rf{functnl} the double-integral representation (cf. with \cite{Dirichlet})
\be
\label{doubint}
\F = {1\over 2}\int_{x_1>x_2} dx_1 dx_2 f''(x_1)f''(x_2) F(x_1-x_2)+ aa^D + \Sigma_0
\ee
expressing it in terms of the perturbative kernel \rf{SWkern} and extremal shape
$f(x)$, solving \rf{exeq}. The (time-dependent) constant $\Sigma_0$ arises in \rf{doubint}
due to constraint \rf{normlog} and appears to be the constant part of the first primitive
of the function \rf{Sfun}.

\section{Dispersionless Toda chain}

In the case of a single cut let us present the double cover of the
$z$-plane $y^2=(z-x^+)(z-x^-)$ in the form
\be\label{u1curve}
z=v+\Lambda\left(w+{1\over w}\right)
\ee
with $x^\pm = v\pm 2\Lambda$ and
\be\label{yu1}
y^2=(z-v)^2-4\Lambda^2
\ee
On the double cover \rf{u1curve}, which is in the
case of single cut just $\mathbb{P}^1$ with two marked points $P_\pm$, with
$z(P_\pm)=\infty$, $w^{\pm 1}(P_\pm)=\infty$, formula \rf{Sfun} defines a function
with a logarithmic cut and asymptotic behavior \rf{sasympt}, odd under the
involution $w\leftrightarrow {1\over w}$ of the curve \rf{u1curve}. In terms
of the uniformizing variable $w$ one can globally write
\be
\label{Sw}
S = -2\left(v+\Lambda\left(w+{1\over w}\right)\right)\log w
-2\Lambda(\log\Lambda-1)\left(w-{1\over w}\right)+
\sum_{k>0} t_k\Omega_k(w) + 2a\log w
\ee
where
\be
\label{Ow}
\Omega_k (w) = z^k_+-z^k_-,  \ \ \ k>0
\ee
are the Laurent polynomials, odd under $w\leftrightarrow{1\over w}$. The first
term in \rf{Sw} comes from the Legendre transform of the Seiberg-Witten differential
$d\Sigma \sim z{dw\over w}$.

The canonical Toda chain times are defined by the coefficients
at the singular terms in \rf{sasympt}
\be
\label{t0res}
t_0 = \res_{P_+} dS = - \res_{P_-} dS = 2a
\ee
and
\be
\label{tP}
t_k = {1\over k}\ \res_{P_+} z^{-k}dS =
- {1\over k}\ \res_{P_-} z^{-k}dS,\ \ \ k>0
\ee
From the expansion \rf{sasympt}
it also immediately follows, that
\be
\label{tPd}
{\d\F\over \d t_k} =  \ha \res_{P_+} z^{k}dS
= - \ha \res_{P_-} z^{k}dS,\ \ \ k>0
\ee
Formulas \rf{t0res}, \rf{tP}, \rf{tPd} together with \rf{aad}
identify the generating function \rf{functnl} with the logarithm of quasiclassical
tau-function, being here, in the case of a single cut, a tau-function of dispersionless Toda
chain hierarchy.

The consistency condition for
\rf{tPd} is ensured by the symmetricity of second derivatives
\be
\label{sysi}
{\d^2\F\over \d t_n\d t_k} = \ha \res_{P_+} (z^k d\Omega_n)
\ee
where the time derivatives of \rf{sasympt}
\be
\label{Oz}
\Omega_0 = {\d S\over\d a}\ \stackreb{z\to P_\pm}{=}\ \pm\left(2\log z
- {\d^2 \F\over\d a^2} - 2\sum_{n>0}
{\d^2 \F\over\d a\d t_n}{1\over nz^n}\right)
\\
\Omega_k = {\d S\over\d t_k}\ \stackreb{z\to P_\pm}{=}\ \pm\left(z^k
- {\d^2 \F\over\d a\d t_k} - 2\sum_{n>0}
{\d^2 \F\over\d t_k\d t_n}{1\over nz^n}\right),\ \ \ k>0
\ee
form a basis of meromorphic functions with poles at the points $P_\pm$, with
$z(P_\pm)=\infty$. All time-derivatives here are taken at constant $z$.

Expansion \rf{Oz} of the Hamiltonian functions \rf{Ow} expresses the second
derivatives of $\F$ in terms of the coefficients of the equation of the curve \rf{u1curve}, e.g.
\be
\label{Omexp}
\Omega_0\ \stackreb{z\to\infty}{=}\ 2\log z - 2\log\Lambda - {2v\over z}
- {2\Lambda^2+v^2\over z^2} + \ldots
\\
\Omega_1\ \stackreb{z\to\infty}{=}\ z - v - {2\Lambda^2\over z} - {2v\Lambda^2\over z^2}
+ \ldots
\\
\Omega_2\ \stackreb{z\to\infty}{=}\ z^2 - (v^2+2\Lambda^2) - {4v\Lambda^2\over z} -
{2\Lambda^2(\Lambda^2+2v^2)\over z^2} + \ldots
\ee
Comparison of the coefficients in \rf{Omexp} gives, in particular,
\be
\label{F2der}
{\d^2\F\over \d a^2} = \log\Lambda^2,
\ \ \ \
{\d^2\F\over \d a\d t_1} = v
\ee
and
\be\label{todaeq}
{\d^2 \F \over \d t_1^2} =\Lambda^2=\exp {\d^2 \F \over \d a^2}
\ee
which becomes the long-wave limit of the Toda chain equations
after an extra derivative with respect to $a$ is taken
\be\label{todaequ}
{{{\d}^2 a^D}\over {{\d t_{1}^2}}} \ = \ {\d \over {\d a}}
\exp {\d a^D \over \d a}
\ee
for the Toda co-ordinate $a^{D} = {{\d\F}\over{\d a}}$. Substituting expansions
\rf{Omexp} into \rf{sysi}, one gets the expressions for the so called contact terms \cite{LNS}
in the $U(1)$ case, which are the polynomials of
a single variable $v$ with $\Lambda$-dependent coefficients.

One can now find the dependence of the coefficients of the curve \rf{u1curve}
on the deformation parameters ${\bf t}$ of the microscopic theory by
requiring $dS=0$ at the ramification points, where $dz=0$. This condition
avoids from arising of extra singularities at the branch points in the variation of
$dS$ w.r.t. moduli of the curve. Equation
\be
\label{dz}
{dz\over d\log w} = \Lambda\left(w-{1\over w}\right) =0
\ee
gives $w = \pm 1$, where now
\be
\label{eqscu}
\left.{dS\over d\log w}\right|_{w =\pm 1} =
\sum_{k>0}t_k \left.{d\Omega_k\over d\log w}\right|_{w =\pm 1}
+ 2a-2v \mp 4\Lambda\log\Lambda =0
\ee
If $t_k=0$ for $k>1$, solution to \rf{eqscu} immediately gives
\be
\label{t01}
v=a, \ \ \ \Lambda^2=e^{t_1}
\ee
and the prepotential
\be
\label{Fsphs}
\F = \ha aa^D + \ha\res_{P_+} \left( z dS\right) -{a^2\over 2} =
\ha a^2t_1+e^{t_1}
\ee
Adding nonvanishing $t_2$, one finds
\be
\label{t012}
v=a -{1\over 2t_2}{\bf L}\left(-4t_2^2e^{t_1+2t_2a}\right)
\\
\log\Lambda^2 = t_1+2t_2a - {\bf L}\left(-4t_2^2e^{t_1+2t_2a}\right)
\ee
where the Lambert function ${\bf L}(t)$ is defined by an expansion
\be
{\bf L}(t) = \sum_{n=1}^{\infty} {(-n)^{n-1} t^{n}\over n!}  =
t - t^2 + {3\over 2} t^{3} - {8 \over 3} t^{4} + \ldots
\ee
and satisfies to the functional equation
\be
\label{Lambert}
{\bf L}(t) e^{{\bf L}(t)}= t
\ee
Hence, for the prepotential with $t_1,t_2\neq 0$ one gets
\be
\label{F}
\F = \ha\left( a{\d\F\over\d a} +
\sum_{k>1}(1-k)t_k {\d\F\over\d t_k}\right) + {\d\F\over\d t_1} -{a^2\over 2} =
\\
=\ha aa^D + {1\over 4}\sum_{k>1}(1-k)t_k  \res_{P_+} \left( z^k dS\right) +
\ha\res_{P_+} \left( z dS\right) -{a^2\over 2} =
\\
\stackreb{t_k=0,k>2}{=}\
\ha aa^D + \ha\res_{P_+} \left( z dS\right)
- {1\over 4} t_2 \res_{P_+} \left( z^2 dS\right) -{a^2\over 2}
\ee
wherefrom the instanton expansion can be computed
(which can be strictly got as an expansion in parameter $q$ after
$t_1\to t_1+\ha\log q$)
\be
\label{F1inst}
\F = {\bf t}(a)
+ e^{{\bf t}''(a)} +
 t_2^2\ e^{2{\bf t}''(a)}
+ {8\over 3}t_2^4\ e^{3{\bf t}''(a)} +
{32 \over 3}t_2^6\ e^{4{\bf t}''(a)} +
\\
+{160 \over 3}t_2^8\ e^{5{\bf t}''(a)}
+ {1536\over 5}t_2^{10}\ e^{6{\bf t}''(a)}
+\ldots =
{\bf t}(a) + S(a) + {1\over 4}S(a)S''(a) + \ldots
\ee
with $S(a)=\exp {\bf t}''(a)$. Expansion \rf{F1inst} directly corresponds
to summing over connected tree diagrams in bosonic model, presented at fig.~\ref{fi:bks}.

It is also easy to compute the explicit form of the extremal shape for
nonvanishing $t_1, t_2$, which reads
\be
\label{shape2}
f^{\prime}(x) =
{2\over {\pi}} \left( {\rm arcsin}\left( {x - v \over 2{\Lambda}} \right) + t_{2}
\sqrt{4 {\Lambda}^{2}-(x-v)^2} \right) \\
v - 2 \Lambda \leq x \leq v + 2 \Lambda
\ee
where $v=v({\bf t})$ and $\Lambda=\Lambda({\bf t})$ are given by \rf{t012}, or obey
\be
\label{rel2}
v-a=2t_2\Lambda^2,\ \ \ \ \log\Lambda^2 = t_1+2t_2v
\ee
Formula
\rf{shape2} is a direct consequence of \rf{dfi} and
directly following from \rf{Sw} upon the relations \rf{rel2}
expression
\be
\label{fit2}
\Phi (w;t_1,t_2,a) = 2t_2y - 2\log w + (t_1+2t_2v-\log\Lambda^2)
{z-v\over y} + 2{a-v+2t_2\Lambda^2\over y} =
\\
\stackreb{\rf{rel2}}{=}\ 2t_2\Lambda \left(w-{1\over w}\right) -
2\log w
\ee
Note, that \rf{shape2} stays that the Vershik-Kerov ``arcsin law'' \cite{VK} for the limiting
shape is deformed by the Wigner semicircle distribution.

If the first three Toda times $t_1,t_2,t_3$ are nonvanishing, instead of \rf{shape2}
one gets
\be
\label{shape3}
f^{\prime}(x) =
{2\over {\pi}} \left( {\rm arcsin}\left( {x - v \over 2{\Lambda}} \right) +
(t_{2}+3t_3v)\sqrt{4 {\Lambda}^{2}-(x-v)^2} +
{3\over 2}t_3(x-v)\sqrt{4 {\Lambda}^{2}-(x-v)^2}\right) \\
v - 2 \Lambda \leq x \leq v + 2 \Lambda
\ee
where $v$ and $\Lambda$ are now subjected to
\be
\label{rel3}
v-a=2(t_2+3t_3v)\Lambda^2
\\
\log\Lambda^2 = t_1+2t_2v +3t_3(v^2+2\Lambda^2)
\ee
Generally we obtain for the limit shape
\be
\label{shapegen}
f^{\prime}(x) =
{2\over {\pi}} \left( {\rm arcsin}\left( {x - v \over 2{\Lambda}} \right) +
\sum_{k>1}t_{k}Q_k(x)\sqrt{4 {\Lambda}^{2}-(x-v)^2} \right) \\
v - 2 \Lambda \leq x \leq v + 2 \Lambda
\ee
where $v$ and $\Lambda$ obey some sort of hodograph equations
$v-a=P_v(v,\Lambda;{\bf t})$, $\log\Lambda^2 = P_\Lambda(v,\Lambda;{\bf t})$
for some polynomials $P_v$ and $P_\Lambda$, whose expansion in Toda times ${\bf t}$
can be easily reconstructed from the presented above formulas of general solution.

\section{Extended non-abelian theory}

In the case of $U(N)$ gauge theory one has to consider solution with $N$ cuts
$\{ {\bf I}_i \}$, $i=1\ldots,N$, which arises after adding to the functional \rf{functnl}
$N$ constraints with the Lagrange multipliers
\be
\F \rightarrow \F + \sum_{i=1}^Na^D_i\left(a_i-\ha\int_{{\bf I}_i} dx\ xf''(x)\right)
\ee
i.e. solution to the integral equation
\be
\label{exeqN}
\sum_{k>0} t_k x^k-\int d{\tilde x} f''({\tilde x})(x-{\tilde x})\left(\log |x-{\tilde x}|-1\right)
=a^D_i,
\ \ \ \ \ x\in {\bf I}_i,\ \ i=1,\ldots,N
\ee
Now it can be expressed in terms of the Abelian integrals on the double cover
\be
\label{dcN}
y^2 = \prod_{i=1}^N(z-x^+_i)(z-x^-_i)
\ee
which is a hyperelliptic curve of genus $g=N-1$.
Define, as before:
\be\label{sfunn}
S (z) = {\bf t}'  ( z)- \int \ dx f''(x) ( z - x)
(\log(z - x) - 1) - a^D
\ee
where the integral is taken over the whole support ${\bf I}=\cup_{i=1}^N{\bf I}_i$,
$a^D={1\over N}\sum_{j=1}^N a^D_j$,
and consider its differential, or
\be
\label{dS}
\Phi(z) = {dS\over dz} = \sum_{k>0}kt_k z^{k-1}-\int dx f''(x)\log(z-x)
\ee
satisfying
\be
\label{ficut}
\Phi(x+i0)+\Phi(x-i0)=0,
\ \ \ \ x\in {\bf I}_i,\ \ \ i=1,\ldots,N
\ee
on each cut, and normalized to
\be
\label{ImfiN}
\Phi(x^+_N)=0,\\
\Phi(x^-_j \pm i0) = \Phi(x^+_{j-1} \pm i0) = \pm 2\pi i (N-j+1),\ \ \ \ j=2,\ldots,N
\\
\Phi(x^-_1)=\pm 2\pi i N
\ee

\bigskip
\mezzo{Vanishing microscopic times}

\noindent
Consider, first, all $t_k=0$ for $k\neq 1$, and define $\Lambda^{2N}=e^{t_1}$.
Now $\Phi={dS\over dz}$ is an Abelian integral on the curve
\rf{dcN} with the asymptotic
\be
\label{fiasysw}
\Phi\
\stackreb{P\to P_\pm}{=}\  \mp 2N\log z \pm 2N\log\Lambda  + O(z^{-1})
\ee
whose jumps are integer-valued due to \rf{ImfiN}, or
$\oint d\Phi \sim 4\pi i\mathbb{Z}$.
It means that the hyperelliptic curve \rf{dcN} can be seen also as an algebraic
Riemann surface for the function
$w=\exp\left(-\Phi/2 \right)$,
satisfying quadratic equation
\be
\label{Todacu}
\Lambda^N\left(w+{1\over w}\right) = P_N(z) = \prod_{i=1}^N (z-v_i)
\ee
since for the two branches $w_+=w$ and $w_-={1\over w}$ one immediately finds
that their product $w_+\cdot w_-$ and sum $w_+ + w_-$ are polynomials of $z$ of given powers (zero
and $N$ correspondingly).

Equivalently, the ends of the cuts in \rf{dcN} are restricted by $N$ constraints in such a way, that
this equation can be rewritten as
\be
\label{Today}
y^2 = P_N(z)^2-4\Lambda^{2N}
\ee
i.e. $\{ x^{\pm}_i \}$ are roots of $P_N(z)\mp 2\Lambda^N=0$, and
\be
\label{yw}
y = \Lambda^N\left(w-{1\over w}\right)
\ee
The generating differential \rf{dS} is now
\be
\label{dSsw}
dS = -2\log w dz = -d(2z\log w) + 2z{dw\over w}
\ee
just the Legendre transform
of the Seiberg-Witten differential $d\Sigma \sim z{dw\over w}$
on the curve \rf{Todacu}, \rf{Today}. It periods
\be
\label{SWper}
a_i = {1\over 2\pi i}\oint_{A_i} z {dw\over w}
\ee
coincide with the Seiberg-Witten integrals and
the only nontrivial residues at infinity give
\be
\res_{P_+}\left( z^{-1}dS\right) = - \res_{P_-}\left( z^{-1}dS\right) = \log\Lambda^{2N}
\\
\res_{P_+}\left(dS\right) = - \res_{P_-}\left(dS\right) = 2\sum_{j=1}^N v_j
\ee
The differential \rf{dSsw} satisfies the condition
\be
\delta dS \sim {\delta w\over w} dz = {\delta P(z)\over y} dz\
\stackreb{\sum_{j=1}^N v_j=0}{=}\ {\rm holomorphic}
\ee
where the variation is taken at constant co-ordinate $z$ and constant scale factor $\Lambda$. Thus,
the integrable system on ``small phase space'' is solved for the
scale $\Lambda^{2N}=e^{t_1}$ and the moduli $v_j$, $j=1,\ldots,N$ of vacua of the $U(N)$ gauge theory,
satisfying the equation $\sum_{j=1}^N v_j = a$
and the transcendental equations for the Seiberg-Witten periods \rf{Aper}.

\bigskip
\mezzo{Nonvanishing microscopic times}

\noindent
When we switch on ``adiabatically'' the higher times \rf{fuv} with $k>1$,
the number of cuts in \rf{exeqN} remains intact, and the differential \rf{dS} can be still
defined on hyperelliptic curve \rf{dcN}. However, now the role of bipole differential
${dw\over w}$ of the third kind is played by
\be
\label{dpsi}
d\Phi = dz\left(\sum_{k>1} k(k-1)t_k z^{k-2}- \int {dx f''(x)\over z-x}\right) =
\\
= \sum_{k>1} k(k-1)t_k d\Omega_{k-1} - 2N d\Omega_0 -4\pi i
\sum_{j=1}^{N-1}d\omega_j
\ee
where $d\omega_i$, $i=1,\ldots,N-1$ are
canonical holomorphic differentials normalized to the $A$-cycles, surrounding
first $N-1$ cuts. The differentials $d\Omega_k$ in \rf{dpsi} are
fixed by their asymptotic at $z\to\infty$
\be
\label{Omas}
d\Omega_k\ \stackreb{z\to\infty}{=}\
\left\{
\begin{array}{c}
  kz^{k-1}dz + O(z^{-2}),\ \ \ \ k>0 \\
  {dz\over z} + O(z^{-2}),\ \ \ \ k=0
\end{array}\right.
\ee
and vanishing $A$-periods
\be
\oint_{A_i}d\Omega_k = 0,\ \ \ k\geq 0,\ \ \ \forall\ i=1,\ldots,N-1
\ee
The nonvanishing periods of $d\Phi$ are fixed by
\be
\label{dpsiperA}
\oint_{A_j} d\Phi = -2\pi i \int_{{\bf I}_j} f''(x) dx =
-2\pi i \left(f'(x_j^+)-f'(x_j^-)\right) = -4\pi i
\ee
which justifies that generating differential $dS$ is still defined modulo $4\pi i dz$.
The only important
difference with the previous case is that integrality of the periods
$\oint d\Phi \sim 4\pi i\mathbb{Z}$, which
was reformulated in terms of an algebraic equation \rf{Todacu} for the theory on ``small phase space'',
remains now a transcendental equation, which cannot be resolved explicitly.

Nevertheless, on the curve \rf{dcN} any odd under hyperelliptic involution differential
can be always presented as
\be
\label{dpsiy}
d\Phi = {s (z)dz\over y}
\ee
where $s(z)$ is a polynomial of power $N+K-2$ in case of nonvanishing
microscopic times $t_1,\ldots,t_K$ up to the $K$-th order. Its higher $K$
coefficients are fixed by leading asymptotic ($t_2,\ldots,t_K$) and the
residue at infinity
\be
\res_{P_\pm} d\Phi = \mp 2N
\ee
and the rest $N-1$ coefficients can be determined from \rf{dpsiperA}. This
fixes completely the differential $d\Phi$ on the curve \rf{dcN} which still remain
to be dependent upon $2N$ (yet arbitrary) branch points $\{ x_j^\pm\}$.

The generating differential $dS$ can be now defined in terms of the Abelian
integral $\Phi(z)$
\be
\label{dSpsi}
dS = \Phi dz = dz\  \int_{z_0}^z d\Phi
\ee
The dependence upon $2N+1$ parameters (the positions of the branch points in \rf{dcN} together
with $z_0$) is constrained by additional to \rf{dpsiperA} vanishing of
the $B$-periods
\be
\label{dpsiperB}
\oint_{B_j} d\Phi = 0,\ \ \ \ j=1,\ldots,N-1
\ee
Integral representation \rf{dS} suggests a natural normalization \rf{ImfiN}, i.e.
\be
\label{normfi}
z_0=x^+_N, \ \ \ \Phi(z_0)=\Phi(x^+_N) = 0
\ee
where $x^+_N$ is the largest among real ramification points $\{ x^\pm_j \}$.
These conditions lead to the following form of expansion of $\Phi(z)$ in the vicinity of
ramification points
\be
\label{psibranch}
 \Phi (z)\ \stackreb{z\to x^\pm_j}{=}\ \Phi(x^\pm_j) + \phi^\pm_j\sqrt{z-x^\pm_j} + \ldots
\\
 \phi^\pm_j = {2s(x^\pm_j)\over\prod'_k\sqrt{(x^\pm_j-x^+_k)(x^\pm_j-x^-_k)}},\ \ \ \ \
 j=1,\ldots,N
\ee
where the constants $\Phi(x^\pm_j)$ are given by \rf{ImfiN}.

The rest $N+1$ parameters are eaten by the periods
\be
\label{Aper}
a_j = \ha\int_{{\bf I}_j} dx xf''(x) =
-{1\over 4\pi i}\oint_{A_j}zd\Phi = {1\over 4\pi i}\oint_{A_j} dS,\
\ \ \ j=1,\ldots,N-1
\ee
together with the residues
\be
\label{respdz}
a = \half\int_{\bf I} dx xf''(x) = -\half\res_{P_+}\left(z d\Phi \right)
\ee
and the "free term" or scaling factor
\be
\label{t1pdz}
t_1 = \res_{P_+}\left(z^{-1}\Phi dz\right)
\ee
Recall once more, that an essential difference with the case of vanishing times is that
for $t_k\neq 0$, the exponent $\exp\left(\Phi\right)$ acquires an essential
singularity at the points
$P_\pm$, and the constraints \rf{dpsiperA}, \rf{dpsiperB} cannot be
resolved algebraically.
The form of the expansion \rf{psibranch} ensures that variation of the generating
differential at constant $z$ w.r.t. moduli
of the curve \rf{dcN}
\be
\label{delShol}
\delta (dS) = \delta \left(\Phi dz\right)=
\\
\stackreb{z\to x^\pm_j}{=}\ {-s(x^\pm_j)\delta x^\pm_j \over\prod'_k\sqrt{(x^\pm_j-x^+_k)
(x^\pm_j-x^-_k)}}{dz\over
\sqrt{z-x^\pm_j}}+\ldots
\simeq {\rm holomorphic}
\ee
is indeed holomorphic.

The Lagrange multipliers
\be
\label{ad}
a^D_i = {\d\F\over \d a_i}
\ee
can be computed by a standard trick. Consider equation \rf{exeqN} for
$i\neq j$ and fix there $x$-variables to be at the ends of corresponding
cuts. Then
\be
\label{adper}
a^D_i-a^D_j = \Re \int_{x_j^+}^{x_i^-} dS = \ha\oint_{B_{ij}} dS
\ee
or
\be
\label{gradF}
{\d\F\over \d a_i} = \ha\oint_{B_i} dS, \ \ \ i=1,\ldots,N-1
\ee
For the time-derivatives of prepotential one can write
\be\label{Ftres}
{\d \F \over \d t_k} = \ha{\rm res}_{P_+} \left( z^k dS \right) =
- {1\over 2(k+1)}{\rm res}_{P_+} \left( z^{k+1}d\Phi \right)
\ee

\section{Quasiclassical hierarchy and explicit results}

From the expansion \rf{dS} in the case of $U(N)$ extended theory it still follows that the
first derivatives of quasiclassical tau-function $\F$ are given by \rf{derF} and
\rf{tPd}, while for the second derivatives one gets \rf{sysi}, or
\be
\label{sysiW}
{\partial^2{\cal F}\over \partial t_n\partial t_m} =
\half{\rm res}_{P_+} (z^m d\Omega_n) =
\half{\rm res}_{{P_+}\otimes {P_+}}\left( z(P)^n z(P')^m W(P,P')\right)
\ee
where we have introduced the bi-differential $W(P,P')=d_Pd_{P'}\log E(P,P')$, with
$E(P,P')$ being the prime form, see \cite{Fay} for the definitions. In
the inverse co-ordinates $z=z(P)$ and $z'=z(P')$ near the point $P^+$ with
$z(P^+)=\infty$ it has expansion
\be
W(z,z') = {dzdz'\over (z-z')^2}+\ldots =
\sum_{k>0} {dz\over z^{k+1}}\ d\Omega_k(z') + \ldots
\ee
The bi-differential $W(P,P')$ can be related with the Szeg\"o kernel \cite{Fay}
\be
\label{fay}
S_e(P,P')S_{-e}(P,P') =  W(P,P') +
 d\omega_i(P)d\omega_j(P')
{\d\over\d T_{ij}}\log \theta_e( 0|T)
\ee
which, for an even characteristics $e \equiv -e$,
has an explicit expression on hyperelliptic curve \rf{dcN}
\be
S_e(z,z') = \frac{U_e(z) +
U_e(z')}{2\sqrt{U_e(z)U_e(z')}} \frac{\sqrt{dz dz'}}{z-z'}
\label{Sz}
\ee
with
\be
\label{Ue}
U_e(z) = \sqrt{\prod_{j =1}^{N} \frac{z -
x_{e^+_j}}{z - x_{e^-_j}}}
\ee
Here $x_{e^\pm_j}$ is partition of the ramification points of \rf{dcN} into two sets,
corresponding to a characteristic $e$. For example, on a small phase
space, when \rf{dcN} turns into the Seiberg-Witten curve \rf{Todacu}, there is a distinguished
partition $e=E$, corresponding to an even characteristics with
\be\label{UE}
U_{E} (z) = \sqrt{P(z)-2\Lambda^N\over P(z)+2\Lambda^N}
\ee
Substituting \rf{fay}, \rf{Sz} into \rf{sysiW} gives
\be
\label{F2}
{\d^2\F\over \d t_n\d t_m} = \half\res_{{P_+}\otimes {P_+}}\left( z(P)^m z(P')^n S_e(P,P')^2\right) - \\
- \half\res_{P_+} \left(z^n d\omega_i\right)\cdot
\res_{P_+} \left(z^m d\omega_j\right){\d\over\d T_{ij}}\log \theta_e( 0|T) =
\\
= P_{nm}^{(e)}(x_{e^\pm_j}) - 2{\partial^2 \F \over\partial a_i\partial t_n}{\partial^2
\F\over\partial a_j\partial t_m}{\d\over\d T_{ij}}\log \theta_e( 0|T)
\ee
where for the "contact polynomials" one gets from \rf{Sz}
\be\label{poluC}
P_{nm}^{(e)}(x_{e^\pm_j}) = {1\over 4} {\rm res}_{{P_+}\otimes {P_+}}
\left({z^kz'^n\over (z-z')^2}\left(1+{U_e(z)\over 2U_e(z')}+
{U_e(z')\over 2U_e(z)}\right)dzdz'\right)
\ee
If calculated in the vicinity of the small phase space
and for the particular choice of characteristic \rf{UE},
residues \rf{poluC} vanish for $n,m<N$, and one gets exactly the conjectured in \cite{LNS}
formula
\be\label{rgsw}
{\partial^2{\cal F}\over \partial t_n\partial t_m}
=  - \ha{\partial u_{n+1}\over\partial a_i}{\partial u_{m+1}\over\partial a_j}\
{\partial\over \partial T_{ij}}\log\ \theta_E( 0|T), \ \ \ \ \ \  n,m<N
\ee
with
\be\label{ucond}
u_n = 2{\d \F\over \d t_{n-1}} =
{1\over n} {\rm res}_{P_+}\left(z^n {P_N'dz\over y}\right) =
{1\over n}\sum_{l=1}^N v_l^n = {1\over n}\langle\Tr\ \Phi^n\rangle
\ee
Equation \rf{rgsw} is a particular case of the generalized dispersionless Hirota relations
for the Toda lattice, derived in \cite{Dirichlet}.

Let us point out, that this derivation of the renormalization group equation \rf{rgsw} in
\cite{MN} is almost identical to developed previously
in \cite{gmmm} for another version of extended Seiberg-Witten theory, which can be
defined by generating differential
\be
\label{dSgmmm}
d{\tilde S} = \sum_{k>0} T_k d{\tilde\Omega}_k = \sum_{k>0} T_k P(z)^{k/N}_+{dw\over w}
\ee
directly on the Seiberg-Witten curve \rf{Todacu} (whose form remained
intact by higher flows $\{ T_k \}$, in contrast
to the quasiclassical hierarchy, determined by \rf{dSpsi}),
and all derivatives were taken at constant $w$.
For example, if $N=1$ and only $T_0$, $T_1$ do not vanish
with $d{\tilde\Omega}_1=d\Omega_1=(z-v){dw\over w}$, one gets
\be
\label{cuT}
{\d\over\d T_1}\left(z-v\right) = {\d\Lambda\over\d T_1}\left(w+{1\over w}\right)
\ee
and
\be
\label{derlT}
{\d d{\tilde S}\over\d T_1} = d\Omega_1 + T_1{\d\Lambda\over\d T_1}\left(w+{1\over w}\right){dw\over w}
= \left(1+{T_1\over\Lambda}{\d\Lambda\over\d T_1}\right)d\Omega_1
\ee
which means, in particular, that the scale factor $\Lambda \sim T_1$ linearly depends on the first time,
in contrast to the exponential dependence in formula \rf{t01}. Indeed,
taking derivatives of \rf{Sw} at constant $z$, instead of \rf{cuT} one gets
\be
\label{cut}
{\d v\over \d t_1} + {\d\Lambda\over \d t_1}\left(w+{1\over w}\right) +
\Lambda\left(w-{1\over w}\right){\d\log w\over \d t_1} = 0
\ee
and therefore
\be
\label{derlt}
{\d S\over \d t_1} = {\d\over \d t_1}\left(t_1\Omega_1 + 2a\log w -
2z\log w
-2\Lambda(\log\Lambda-1)\left(w-{1\over w}\right)\right) =
 \Lambda\left(w-{1\over w}\right) +
\\
+ {\d\Lambda\over\d t_1}\left(t_1-2\log\Lambda\right)\left(w-{1\over w}\right)+
{\d\log w\over \d t_1}\left((t_1-2\log\Lambda)\Lambda\left(w+{1\over w}\right) +2a-2v \right)
\ee
i.e. formulas \rf{Ow}, \rf{Oz} are provided directly by \rf{t01}.

The choice of extension \rf{dSgmmm} in \cite{gmmm} was motivated rather by technical reasons: preserving
the form of the Seiberg-Witten curve \rf{Todacu} with deforming only the generating differential, moreover
that the latter remained single-valued even in the deformed theory. We see, however, that the old choice
is not
consistent with the microscopic instanton theory \rf{fuv}, \rf{zuone}, \rf{zunntre}, basically
since the appropriate co-ordinate for the quasiclassical hierarchy is $z$,
coming from the scalar field $\Phi$ of the vector multiplet of \N2 supersymmetric gauge theory.
However, the
corresponding quasiclassical hierarchy is defined even more implicitly, due to
the highly transcendental ingredient $\oint d\Phi \sim 4\pi i\mathbb{Z}$ for the second kind
(not for the third kind) Abelian differential, and one
needs to apply special efforts to extract explicit results.

\bigskip
\mezzo{Instanton expansion in the extended theory}

\noindent
The instantonic expansion $\F = \sum_{k\geq 0}\F_k$ in the non-Abelian theory starts
with the perturbative prepotential
\be
\label{prepert}
\F_0 = \sum_{j=1}^N{\bf t}(a_j)+
\sum_{i\neq j}F(a_i-a_j)
\ee
defined entirely in terms of the functions \rf{fuv} and \rf{SWkern}. It is
totally characterized by degenerate differential \rf{dpsiy}
\be
\label{dPhi0}
d\Phi_0 = {\bf t}'''(z)dz - 2{dP_N(z)\over P_N(z)}
= {\bf t}'''(z)dz - 2\sum_{j=1}^N{dz\over z-v_j}
\ee
(which does not depend on higher times),
and the coefficients of the polynomial $P_N(z)$ in \rf{Todacu}, \rf{dPhi0}
coincide with the perturbative values of the Seiberg-Witten periods
\be
\label{avpert}
a_i = - \half\res_{v_i} zd\Phi_0 = v_i
\ee
The perturbative generating differential is $dS_0=\Phi_0 dz$, with
\be
\label{Phi0}
\Phi_0 = {\bf t}''(z) - 2\sum_{j=1}^N\log\left( z-v_j\right)
\ee
and satisfies
\be
{\d dS_0\over \d a_j} = 2{dz\over z-v_j},\ \ \ \ j=1,\ldots,N
\\
{\d dS_0\over \d t_k} =  kz^{k-1}dz,\ \ \ \  k>0
\ee
what gives rise to
\be
S_0(z) = {\bf t}'(z) - 2\sum_{j=1}^N(z-v_j)\left(\log(x-v_j)-1\right)
\ee
Equations
\be
\label{SWpertS}
a^D_j = {\d \F_0\over\d a_j} = 2S_0(a_j)
\ee
completely determine \rf{prepert},
since on this stage one makes no difference between $v_j$ and $a_j$.

Moreover, vanishing of the $B$-periods \rf{dpsiperB} of the differential \rf{dPhi0}
\be
\int_{x_i^+}^{x_j^-} d\Phi_0 =0
\ee
where $x_j^\pm = a_j \pm \sqrt{qS_j}+O(q^2)$ are positions of the branching points of
the curve \rf{dcN} in the vicinity of perturbative rational curve, immediately gives
the deviations
\be
\label{Si}
S_i \sim {e^{{\bf t}''(a_i)}\over\prod_{j\neq i}(a_i-a_j)^2},\ \ \
i=1,\ldots,N
\ee
where the numeric coefficient is fixed from comparison with the Seiberg-Witten curve
\rf{Today} on a small phase space. The instantonic expansion, similarly to that of
the $U(1)$ theory \rf{F1inst}, can be developed in terms of the functions \rf{Si}
and their derivatives.
For example, in \cite{MN} we have checked, that
\be
\label{1inst}
\F_1=\sum_l S_l
\ee
for $U(2)$ gauge group and the only nonvanishing $t_1$, $t_2$, using instantonic expansions
of the equations \rf{dcN}, \rf{dpsiy} and \rf{Ftres}.

\bigskip
\mezzo{Elliptic uniformization for the $U(2)$ theory}

\noindent
In the case of $U(1)$ theory the problem was solved explicitly by construction of the function
\rf{Sw} due to
explicit uniformization of the rational curve \rf{u1curve} in terms of ``global''
spectral parameter $w$. This is hardly possible for generic non-Abelian theory with
the hyperelliptic curve \rf{dcN} of genus $g=N-1$, but in the next to rational case with $N=2$
\be
\label{ellcu}
y^2=\prod_{i=1,2}(z-x^+_i) \equiv  \prod_{i=1}^4(z-x_i) \equiv R(z)
\ee
it is an elliptic curve, and therefore can be uniformized using, for example, the
Weierstrass functions
\be
\label{unif}
z = z_0 + {{R'(x_0)/ 4}\over \wp(\xi)-{R''(x_0)/ 24}} =
z_0 + {\wp'(\xi_0)\over\wp(\xi)-\wp(\xi_0)} =
\\
= z_0 + \zeta(\xi-\xi_0)-\zeta(\xi+\xi_0)+2\zeta(\xi_0)
\\
{dz\over d\xi} = - {\wp'(\xi_0)\wp'(\xi)\over \left(\wp(\xi)-\wp(\xi_0)\right)^2} = y
\ee
where it was convenient to take
\be\label{x04}
z_0=x_4
\\
R'(z_0)=x_{43}x_{42}x_{41} = 4\wp'(\xi_0)
\\
R''(z_0) = 2\left( x_{41}x_{42}+x_{41}x_{43}+x_{42}x_{43}\right) = 24\wp (\xi_0)
\ee
The Abelian integral for $\Phi(z)$ can be now performed in terms of the elliptic
functions. Take again for simplicity all $t_k=0$, if $k>2$. Then for the differential \rf{dpsiy}
one has
\be
\label{dfiel}
d\Phi = {2t_2 z^2 + s_1 z + s_0\over y} dz\ \stackreb{z\to\infty}{=}\ \pm \left(2t_2 dz -
4{dz\over z} - 2a{dz\over z^2} + \ldots\right)
\ee
and, as was promissed before, this asymptotic fixes the coefficients
\be
\label{ab}
s_1 = - 4 - t_2\sum_{i=1}^4x_i
\\
s_0 = -2a + 2\sum_{i=1}^4x_i - {t_2\over 4}\sum_{i=1}^4x_i^2 + {t_2\over 2}\sum_{i<j}x_ix_j
\ee
and completely determines here the differential \rf{dfiel} in terms of the curve \rf{ellcu}.
For the elliptic integral in \rf{dSpsi} one can now write
\be
\label{fiel}
\Phi = \int_{z_0}^z d\Phi = 2t_2\left(\zeta(\xi+\xi_0)+\zeta(\xi-\xi_0)\right)
 + \Phi_1\log{\sigma(\xi_0-\xi)\over\sigma(\xi_0+\xi)} + \Phi_2 \xi
\ee
The constants $\Phi_{1,2}$ are easily recovered from comparison of the expansion of
\be
\label{dfielxi}
{d\Phi\over d\xi} = -2t_2\left(\wp(\xi-\xi_0)+\wp(\xi+\xi_0)\right)+
\Phi_1\left(\zeta(\xi-\xi_0)-\zeta(\xi+\xi_0)\right)+\Phi_2
\ee
at $\xi\to\pm\xi_0$ with \rf{dfiel} upon \rf{unif}.
The jumps of a multivalued Abelian integral \rf{fiel} on the elliptic curve \rf{ellcu},
are further constrained to integers by \rf{dpsiperA} and \rf{dpsiperB}, which can
be now rewritten in the form of transcendental constraints for the parameters of the
Weierstrass functions.

\section{Conclusion}

We have discussed in these notes the main properties of the quasiclassical hierarchy, underlying the
Seiberg-Witten theory, which was derived in \cite{MN} directly from the microscopic setup and
instanton counting. Most of the progress was achieved due to existence of the ``oversimplified''
$U(1)$ example, naively completely trivial from the point of view of the Seiberg-Witten theory.
However, even in this case the partition function of the deformed instantonic theory
becomes a nontrivial function on the large phase space, being the tau-function
of dispersionless Toda chain, and providing a direct link to the theory of the
Gromov-Witten classes. The dual Seiberg-Witten period (``the monopole mass'') satisfies the
long wave limit of the equation of motion in Toda chain, as a function of ``W-boson mass'' and
the (logarithm of) the scale factor.
Much less transparent non-Abelian quasiclassical solution is nevertheless
constructed using standard machinery on higher genus Riemann surfaces. It is also essential, that
switching on higher times deform the Seiberg-Witten curve.

The main issue now is what is going beyond the quasiclassical limit. Could at least the ``simple''
$U(1)$ problem be solved exactly in all orders of string coupling in more or less explicit form?
It is also necessary to stress, that the free fermion (or boson) matrix
elements up to now were considered as formal series in the higher times, except for $t_1\sim\log\Lambda$.
Their knowledge as exact functions at least of $t_2$ could provide us an interesting information
about physically different ``phases" of the model. Another interesting and yet unsolved
problem for the extended theory is switching on the matter by extrapolating the higher
flows to $t_k \sim {1\over k}\sum_{A=1}^{N_f}m_A^{-k}$, what corresponds hypothetically
to the theory with $N_f$ fundamental hypermultiplets with corresponding masses. We hope to return to
these problems elsewhere.

\bigskip
\mezzo{Acknowledgements}

\noindent
I am grateful to A.~Alexandrov, H.~Braden, B.~Dubrovin, I.~Krichever, A.~Losev, V.~Losyakov, A.~Mironov,
A.~Morozov and, especially, to N.~Nekrasov and S.~Kharchev for the very useful discussions.

The work was partially supported by Federal Nuclear Energy Agency, the RFBR grant 05-02-17451,
the grant for support of Scientific Schools 4401.2006.2,
INTAS grant 05-1000008-7865, the project ANR-05-BLAN-0029-01, the
NWO-RFBR program 047.017.2004.015, the Russian-Italian RFBR program 06-01-92059-CE, and by the
Dynasty foundation.

\end{document}